\documentclass[aps,prl,twocolumn,showpacs,showkeys]{revtex4}
\usepackage{graphicx,amsmath}
\usepackage{amssymb}
\usepackage{dcolumn}
\usepackage{bm}

\begin{document}

\title{Descriptive Thermodynamics}

\author{David Ford}
\author{Steven Huntsman}%
 \email{schuntsm@nps.edu}
\affiliation{%
Physics Department, Naval Postgraduate School\\
833 Dyer Road, Monterey, California
}%

\date{\today}

\begin{abstract}
Thermodynamics (in concert with its sister discipline, statistical physics) can be regarded as a data reduction scheme based on partitioning a total system into a subsystem and a bath that weakly interact with each other. Whereas conventionally, the systems investigated require this form of data reduction in order to facilitate prediction, a different problem also occurs, in the context of communication networks, markets, etc. Such ``empirically accessible'' systems typically overwhelm observers with the sort of information that in the case of (say) a gas is effectively unobtainable. What is required for such complex interacting systems is not prediction (this may be impossible when humans besides the observer are responsible for the interactions) but rather, {\sl description} as a route to understanding. Still, the need for a thermodynamical data reduction scheme remains. In this paper, we show how an empirical temperature can be computed for finite, empirically accessible systems, and further outline how this construction allows the age-old science of thermodynamics to be fruitfully applied to them. 
\end{abstract}

\maketitle

\section{ Motivation: the passive demon}

In the following we outline a program for an alternative approach to thermodynamics and statistical physics for empirically accessible systems. Here, ``empirically accessible'' is taken to indicate that the microscopic details of the system-bath interaction are available not merely in principle but in actual fact. (We might properly speak of an empirically accessible membrane, but this distinction also will not bother us here.)
The motivating (but by no means only) example is that of digital communication networks (see, e.g., \cite{Bur},  \cite{TTS}), where a particular network interacting with a larger network (e.g., the Internet), constitutes the system-bath pair, and the data-exchange interactions are mediated by routers, switches, etc. that collectively act as the membrane between the system and bath. In such a case a system administrator or researcher has access to everything that is happening at the membrane on a microscopic level: i.e., the individual packets being routed through the membrane are available for inspection by a passive demon. 

Before going further let us consider a more traditional example, that of a thermally conducting container filled with gaseous helium immersed in a bath of gaseous argon. Here the membrane is the container itself, and a passive demon monitors it, keeping track of all the impacts of the atoms on the two surfaces of the container, and performing the relevant averages. It can be taken for granted that, when the substances are in thermal equilibrium, this demon is able to compute the temperature \cite{Car}. The question therefore arises as to what makes the latter case any different in principle than the former, or some other one like a granular medium \cite{Col}.  

We claim that there is no essential difference. If, like our notional passive demon, we can obtain mesoscopic temporal and ``spatial'' averages as an observer at the membrane, and also keep records pointing to any information discarded in the averaging process, then we ought to be able to do thermodynamics from the bottom up, rather than from the top down. That is, an empirically accessible system affords us the opportunity to do thermodynamics in order to {\sl describe} the system. This is a stark departure from the usual cases, in which systems are empirically inaccessible and thermodynamics is used as a way of predicting the behavior of a system whose microdynamical equations are known, but whose microstate is effectively unknowable. 

In the course of this paper we will demonstrate how empirical temperatures can be calculated for any finite system whose trajectory through state space is available. As a corollary, we will obtain severe algebraic and topological restrictions for any empirical temperature defined as a function of occupation times. We will also describe an essentially universal mesoscopic data format for empirically accessible systems, enabling the explicit computation of temperatures. This in turn enables the entire thermodynamical/statistical-physical apparatus to be brought to bear on real-world problems. With that in mind, we will finally sketch an application.

 \section{Isotherms}

Most of this section derives from \cite{Fo2}. We begin with some preliminaries. For the purposes of clarity and concreteness, we will be fairly specific in some of the details, but much (if not most) of the discussion to follow admits generalization/variation on the minor points. We restrict the discussion to systems with a fixed, finite state space, where the states are labeled $\{1,2,\cdots, n\} $ for convenience and have variable (but as yet undetermined) energies $\{E_1,E_2,\cdots, E_n\} $. 
The trajectory of the system through the state space evolves in continuous time starting at $t = 0$, and (again, for clarity and concreteness) the trajectory can be decomposed into disjoint epochs according to (for example) the following protocol.

The $(n+1)$th epoch starts at $t =  t_{\infty}(s_0 = 0) + ... + t_{\infty}(s_n)$ and ends as soon as every state has been visited at least once during the epoch: its duration$-$the {\sl Carlson depth of the ith epoch}$-$is denoted $t_{\infty}(s_n)$. 

In this way we will effect a mesoscopic moving time average over a fixed ``length'' scale (reflected in the cardinality of the state space). During the ith epoch, the state occupation times are taken to be $\{t_1(s_i), â\cdots, t_n(s_i)\}$, so that 
\begin{equation}
\sum_{k=1}^{n} t_k(s_i) = t_{\infty}(s_i).
\end{equation}

A second preliminary point deserves mention. Although the arithmetical structure of statistical physics allows the arbitrary numerical evolution of state energies, in a canonically distributed system we are constrained by the physics to interpret the underlying dynamics as simultaneous changes to the energy levels and to the bath temperature. Another way of saying this is that temperature parametrizes the total bath energy. As a result, we are free to stipulate that the state energies always satisfy
\begin{equation}
\sum_{k=1}^{n} E_k = 0.
\end{equation}
This is just re-attributing homogeneous fluctuations in the state energies to temperature fluctuations.
	With this stipulation in mind, we introduce an empirical temperature parameter $\Theta$, and two sets of coordinate systems: experimentalist's and theorist's. See table \ref{coords}.

\begin{table}
\caption{\label{coords} Notations employed for the $n-$dimensional
spaces used in this work. Both the vector notations and their coordinates
are shown. }
\begin{ruledtabular}
\begin{tabular}{lcr}
Experimentalist&Theorist\\
\hline
$\phantom{XXX} \bm{t}(s_i) $ & $\bm{H}(s_i)$\\
$ \{t_1(s_i),\cdots, t_n(s_i)\}$&  $\{E_1(s_i),\cdots, E_{n-1}(s_i),\Theta(s_i)\}$\\
\end{tabular}
\end{ruledtabular}
\end{table}
It is natural to ask how to transform from one set of coordinates to the other. The transformation in one direction, from $\bm{H}-$ to $\bm{ t}-$space, is the conventional aim of statistical physics. The other direction is far less explored, and negotiating it will allow us to bring the tools of thermodynamics and statistical physics to bear on a host of real-world problems. Without any constraints (e.g., the internal energy of the system is not fixed), a Boltzmann-Gibbs {\sl Ansatz} for the state occupation probabilities is underdetermined, i.e., the assignment
\begin{equation}
\frac{t_k(s_i)}{t_{\infty}(s_i)} = p_k^{(t)} \equiv p_k^{(H)}(s_i) = \frac{e^\frac{-E_k(s_i)}{\Theta(s_i)}}{Z(s_i)}
\end{equation}

still does not uniquely specify a point in $\bm{H}-$spaceâ but only a ray.

Suppose that the state probabilities change slightly, say from one epoch to the next: we have the innocuous equation 
\begin{equation}
\Delta p_k^{(H)}(s_i,s_{i+1}) \equiv  \Delta p_k^{(t)}(s_i,s_{i+1})
\end{equation}
\noindent which embodies the essence of the {\sl Matched Invariants Principle} (MIP): probability shifts agree in both coordinate systems.

In what follows, we can dispense with the specification of particular epochs. The MIP implies that the total differential of each of the state probabilities can be written in two ways:
\begin{eqnarray}\label{6}
\langle  \nabla^t p_k^{(t)} \mid d\bm{t} \rangle = & dp^t_k ; \; \langle  \nabla^H p_k^{(H)} \mid d\bm{H} \rangle =  & dp^H_k 
\end{eqnarray}

Now for each set of gradients, only $n-1$ are linearly independent. In particular, the sets of gradients each span tangent spaces to spheres, because probabilities are constant on rays in both $\bm{H}-$ and $\bm{ t}-$ space. 
	Consequently, we can expand the differentials in terms of (say) the first $n-1$ gradients:
\begin{eqnarray}\label{7}
d\bm{H} = & \sum_{k=1}^{n-1} \frac{ \nabla^H p_k^{(H)}}{\lVert \nabla^H p_k^{(H)} \rVert} dr^H_k; \;  d\bm{t} =  & \sum_{k=1}^{n-1} \frac{ \nabla^t p_k^{(t)}}{\lVert \nabla^t p_k^{(t)} \rVert} dr^t_k 
\end{eqnarray}

Here we have introduced the differentials of new affine parameters in the $dr$ quantities. They are defined so that (\ref{7}) holds. Because the first $n-1$ gradients are linearly independent and span tangent spaces to spheres, these affine differentials are well-defined quantities.

Combining (\ref{6}) and (\ref{7}) immediately yields the thermodynamic evolution equation (TEE)
\begin{eqnarray}
d\bm{r}^H = ( \hat{M}^{(H)})^{-1} \hat{M}^{(t)} d\bm{r}^t = J^{(t,H)} d\bm{r}^t
\end{eqnarray}

\noindent where we have introduced the matrices
\begin{eqnarray}
M_{jk}^{(t)} :=\frac{ \langle  \nabla^t p_k^{(t)} \mid \nabla^t p_j^{(t)} \rangle}{\lVert \nabla^t p_j^{(t)} \rVert} 
\nonumber \\
M_{jk}^{(H)} :=\frac{ \langle  \nabla^H p_k^{(H)} \mid \nabla^H p_j^{(H)} \rangle}{\lVert \nabla^H p_j^{(H)} \rVert}  
\end{eqnarray}

The TEE is a system of first-order ordinary differential equations, and the standard existence and uniqueness results apply. The integral curves of the TEE in either space are circular arcs transverse to the foliation induced by the isotherms (i.e., the set of level hypersurfaces of constant temperature). This generic transversality condition implies, among other things, that the isotherms are pinched at the origin in $\bm{t}-$space. The Jacobian $J$ simply relates the length parametrizations of these arcs under a change of variables, and we can use this fact to obtain isotherms numerically.

\section{Example: a two-state system} 

	In the case of a two-state system, the preceding considerations suggest working in terms of a single probability and a single energy, i.e., we have that
\begin{eqnarray}
p_1 =  p, \; p_2 = 1 - p; 
\; E_1=E, \; E_2 = -E
\end{eqnarray}
and some routine algebra leads from here to
\begin{eqnarray}
p=  \frac{1}{1+e^{2 \frac{E}{\Theta}}}; \;
 \frac{E}{\Theta} = \frac{1}{2} \log{\frac{1-p}{p}}=\tan{\phi}
 \end{eqnarray}
 
 The probability gradients are a bit more involved to compute:
\begin{eqnarray}
\nabla^{t} p=  -\frac{1}{t_{\infty}}  
 \begin{pmatrix}
  p\\ 
  1-p
   \end{pmatrix};  \;
\nabla^{H} p=  -\frac{p}{\Theta}  
 \begin{pmatrix}
  2 (1 - p)\\ 
  \frac{U-E}{\Theta}
   \end{pmatrix} 
 \end{eqnarray}

The somewhat complicated appearance of the $\bm{H}-$gradient is due to the constraint implied by the two probabilities summing to unity. 
	In this example, the TEE is a scalar ODE, and we find that
\begin{eqnarray}
\frac{dr^{(t)}}{dr^{(H)}}=  \frac{\lVert \nabla^H p^{(H)} \rVert}{\lVert \nabla^t p^{(t)} \rVert} =  \frac{\Theta}{t_{\infty}} \sqrt{ f_{F,2} } 
\end{eqnarray}
 
 where 
 \begin{eqnarray}
 f_{F,2} :=  \frac{ p^{-2} + (1- p)^{-2} }{4 + \log^2{(p^{-1} - 1})}.
 \end{eqnarray}

A geometrical argument (figures 1 and 2) allows for a semi-analytical determination of the isotherms in terms of a bijective {\sl uniform temperature map} (UT map), abusively denoted
 \begin{eqnarray}
\Theta :  \{\bm{t}: \bm{t} = \frac{t_{\infty}}{\sqrt{n} } \,  \hat{\bm{u}}^t \} \rightarrow 
 \{\bm{H}: \bm{H} = \Theta \,  \hat{\bm{u}}^H\}
 \end{eqnarray}

where we introduce the unit {\sl uniform vectors}
\begin{eqnarray}
 \hat{\bm{u}}^t  = \frac{1}{\sqrt{n}} \bm{1};
 \hat{\bm{u}}^H = \hat{e}_{\Theta} = \{0,0,\cdots,1\}.
 \end{eqnarray}
The requirement for bijectivity of the UT map is just a codification of the zeroth law of thermodynamics.

 \begin{figure}[htbp]
\includegraphics[width=60mm,keepaspectratio]{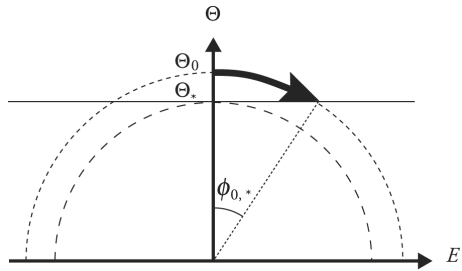}
\caption{   \label{fig:semi} Schematic geometry of the transformation for a two-state system (higher dimensions are not qualitatively any different).}
\end{figure}

\begin{figure}[htbp]
\includegraphics[width=60mm, keepaspectratio]{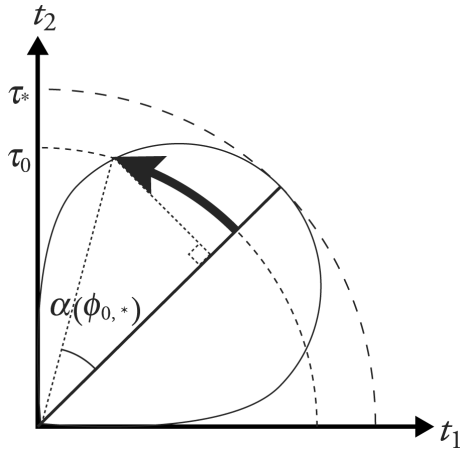}
\caption{ \label{fig:quarter} Schematic of an isotherm in $\bm{t}-$space.}
\end{figure}

 The semi-circles in $\bm{H}-$space (see figure \ref{fig:semi}) map to the corresponding quarter-circles in $\bm{t}-$space (figure \ref{fig:quarter}). Given a uniform temperature map, the concomitant isotherms in $\bm{t}-$space can be identified numerically by integrating the TEE.

	Briefly, suppose that we begin with the corresponding uniform points
\begin{eqnarray}
\bm{t}_0 = \frac{t_{\infty}}{\sqrt{n}} \hat{\bm{u}}^t  \equiv \tau_0 \hat{\bm{u}}^t ; \;
\bm{H}_0 = \Theta(\tau_0) \, \hat{\bm{u}}^H \equiv \Theta_0 \, \hat{\bm{u}}^H. 
 \end{eqnarray}

\noindent and jointly follow the corresponding integral curves according to
\begin{eqnarray}
\int dr^t  = \tau_0 \; \alpha(\phi_{0,*}); \;
\int dr^H  = \Theta_0 \; \phi_{0,*}.
\end{eqnarray}

The angles are defined in accordance with the figures, which are also important aids to the following arguments.
	We have that $\Theta_* = \Theta_0 \; \cos(\phi_{0,*})$, and combining the last few equations, we find that 
\begin{eqnarray}
\alpha(\phi_{0,*}) = \frac{1}{\tau_0}\int_0^{\phi_{0,*}} \frac{dr^t }{dr^H } \; dr^H  
= \frac{\Theta_0}{\tau_0}\int_0^{\phi_{0,*}} \frac{dr^t }{dr^H } \; d\phi \nonumber \\
= \frac{\Theta_0^2}{\tau_0^2} \int_0^{\phi_{0,*}} \frac{\cos{\phi} }{t_{\infty}(\phi) } 
\sqrt{f_{F,2}(\phi)}\; d\phi
\end{eqnarray}

Referring to the geometry of the $\bm{H} \leftrightarrow \bm{t}$ transformation, we see that by integrating along circular arcs this could be solved numerically. However there is not much point in bothering: the differential equation will do just as well for numerical purposes. The point in this exercise is twofold. First, it highlights the geometry governing the limits of integration for the TEE for an arbitrary number of states. Second, it details the need for, and the use of, a UT map in closing the transformation equations.

\section{Parameterization of the the isotherms}
From the preceding considerations it is clear that once a UT map is selected, the isotherms are obtained. Although the detailed geometry of the isotherms depends on the choice of UT map, some physical considerations allow us to immediately single out a particular class of candidate UT maps that all exhibit the same qualitative features.

Let us return momentarily to the specific example of the container of helium immersed in argon. To a first approximation, we can consider the atoms classically with conservative hard-body interactions. Imagine that in this idealization an active demon, through a homogeneous dilatation of the velocities, doubles the kinetic energies of all the argon atoms. The temperature (as the volume averaged kinetic energy of the atoms) that our passive demon computes is also doubled.
 
Now let us assume that the passive demon has a very poor sense of inertia, and that our active demon does something else in addition to the foregoing: he replaces the passive demon's argon atoms with neon atoms of half the mass. The result of this trickery is, of course, that when the passive demon computes kinetic energies and temperature the original values are maintained. This suggests that if the same events happen more quickly, the temperature ought to be higher \cite{Fo3}.

We can see as a result that the temperature ought to approach zero (by bijectivity, monotonically) as the Carlson depth approaches infinity. Conversely, the temperature ought to approach infinity as the Carlson depth approaches zero. A UT map satisfying these requirements will be called topologically admissible (TA), as will the concomitant temperature function on $\bm{t}-$space. Now we have a quite detailed characterization of any topologically admissible temperature function. A caricature is provided by two examples below (Figures 3 and 4).

\begin{figure}[htbp]
\includegraphics[width=75mm, keepaspectratio]{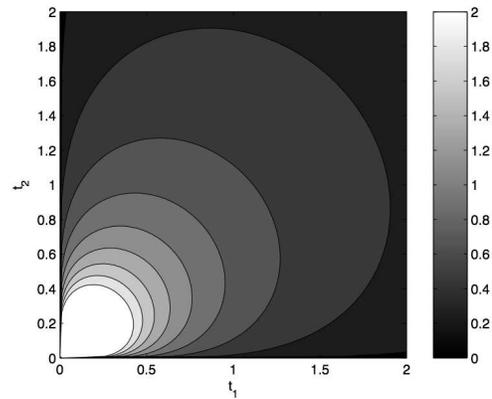}
\caption{   \label{fig:gamm1} Temperature function for a two-state system with the uniform temperature map $(\frac{\sqrt{n}}{t_{\infty}})^{\gamma}$ with 
$\gamma = 1$.  Both this figure and the next one were generated by an analytic formalism due to one of the authors \protect\cite{Fo3} that agrees with the numerical approach discussed here.}
\end{figure}

\begin{figure}[htbp]
\includegraphics[width=75mm, keepaspectratio]{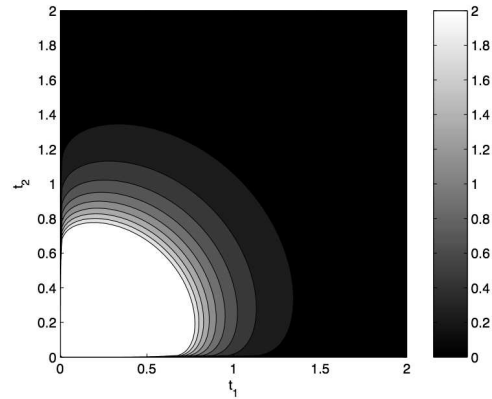}
\caption{   \label{fig:gamma4} Temperature function for a two-state system with the uniform temperature map
$(\frac{\sqrt{n}}{t_{\infty}})^{\gamma}$ with $\gamma = 4$. }
\end{figure}
	It is perhaps natural to wonder if a particular TA UT map (or subclass of TA UT maps) might apply to every system. Although an answer might appear intractable, field-theoretical considerations (e.g., examination of the KMS condition \cite{Par}) appear to suggest a UT map of the form $\Theta = \frac{k}{t_{\infty}}$, where $k$ is some constant. Publication of work in this direction is in progress by one of the authors \cite{Fo3} and will be reported elsewhere, together with a straightforward algebraic derivation of the isotherms in terms of a UT map.

\section{The closed queue format}

	Now that we have an empirical temperature, the issue arises of how to apply it to an empirically accessible system. Towards that end we return once more to the metaphor of the passive demon monitoring abstract transactions across the membrane. 
In the case of our idealized gases, these transactions fall into several classes according to some discretization of the phase space. For instance, consider an argon atom with momentum $\bm{p}$ impinging upon the outside of the container at position $\bm{x}$, where it donates some energy to the container and scatters.

As a simplified energy transfer mechanism we might assume that the donated energy remains relatively localized as a phonon (it does not have time to diffuse much) before being absorbed by a helium atom impinging on the inside. The passive demon can say that a transition of the form
\begin{eqnarray}\label{phonon}
(\bm{p}, \bm{x};Ar) \xrightarrow{phonon} (\acute{\bm{p}}, \acute{\bm{x}}; He)
\end{eqnarray}

\noindent occurred. We might generalize this example somewhat so that both system and bath contain a mixture of all the noble gases.

Having discretized phase space near the membrane, the passive demon can associate triplets as above to discrete indices. The demon might choose to consider other observables too. Then a transition of the form above is recast as
\begin{eqnarray}\label{jth}
j \xrightarrow{phonon} \acute j
\end{eqnarray}
In general, we require that all indices for the system and bath are common: say that there are $B$ of them. The map from the tuples in (\ref{phonon}) to the indices in (\ref{jth}) defines the mesoscopic observables or {\sl bucket labels}.

Our mesoscopic data format is now described by fixing some positive integer $b$ and considering the {\sl bucketspace} or set of configurations of $b$ indistinguishable balls in $B$ buckets:
\begin{eqnarray}
X_{B,b}:= \{ \alpha \in \bm{N}^{B}: \alpha_1 + \cdots+\alpha_B=b \}
\end{eqnarray}

At the start of each epoch, the passive demon initializes a new mini-trajectory through bucketspace as follows: the trajectory begins at a fixed point in bucketspace. Every time a transition of the form above occurs, the passive demon moves a ball from bucket $j$ to bucket $\acute j$ if such a move is possible: otherwise, he takes a ball from bucket $j$ and replaces it. This sort of construction can trace its heritage back to the Ehrenfests \cite{Ehr} and Siegert \cite{Sie}, among others. 

At the end of each epoch, the passive demon reports data like the transition rates and transition probabilities between buckets, the average state, etc. as well as the occupation probabilities $p_{\alpha}$. Using these occupation probabilities, the temperature (and from it, other quantities of interest) can be computed according to the scheme outlined above. More can be done: characterizing equilibria, fluctuations, and symmetries are all the subjects of ongoing work by the authors \cite{Hun}. 

A sketch of how the preceding considerations can be fruitfully applied to real-world problems is in \cite{FoH}.

\begin{acknowledgments}
The authors thank Debbie Huntsman for help in generating figures, and James Luscombe for helpful conversations.
\end{acknowledgments}

\newpage 

\bibliography{FHbib}
\end{document}